\documentclass{article}
\usepackage{amssymb}

\usepackage{makeidx}
\usepackage{graphicx}
\usepackage{amsmath}

\input{tcilatex}

\begin{document}

\title{Stability of mixed Nash equilibria in symmetric quantum games}
\author{A. Iqbal and A. H. Toor \\
%EndAName
Department of Electronics, Quaid-i-Azam University, \\
Islamabad 45320, Pakistan.\\
}
\maketitle

\begin{abstract}
In bi-matrix games the Bishop-Cannings theorem of the classical evolutionary
game theory does not permit pure evolutionarily stable strategies (ESSs)
when a mixed ESS exists. We find the necessary form of two-qubit initial
quantum states when a switch-over to a quantum version of the game also
changes the evolutionary stability of a mixed symmetric Nash equilibrium.
\end{abstract}



\section{Introduction}

Quantum game theory has gained considerable interest recently. In a
pioneering work, Meyer \cite{meyer} presented the idea of playing a quantum
form of a sequential game by unitary manipulation of a qubit. A measurement
of the final quantum state of the qubit gives the payoffs to the players.
Eisert, Wilkens, and Lewenstein \cite{eisert}, while focussing on the
concept of Nash equilibrium (NE)\emph{\ }\cite{Nash}\emph{\ }from
noncooperative game theory, extended the famous game of prisoner's dilemma
to quantum domain. Using a maximally entangled two-qubit initial quantum
state, they showed that the dilemma can be made to disappear when players
have access to a particular set of unitary operators. Also the classical
game can be reproduced as a subset. Later Marinatto and Weber \cite
{marinatto} followed a different approach and studied the game of battle of
sexes in quantum settings, showing that the introduction of entangled
strategies leads to a unique solution of this game. Moreover, they showed
that in their scheme the classical game corresponds to an unentangled
initial quantum state.

An important question in quantum game theory is to draw a comparison with
the corresponding classical version of the game. In classical game theory
there is well developed mathematical formalism to study the evolutionary
dynamics of a population consisting of interacting individuals \cite
{smith,weibull}. It is interesting to investigate and extend this formalism
to quantum domain, and also to compare the predictions of classical and
quantum game-theoretical models of evolution. In other words, how the
established evolutionary concepts of mathematical biology, based on
classical game-theoretical modeling, are modified by the introduction of
Hilbert space? In our earlier papers \cite{iqbal,iqbal1,iqbal2,iqbal3} we
explored the relevance of the concept of evolutionary stability in quantum
game theory. In evolutionary game theory, an \textit{evolutionarily stable
strategy} (ESS) \cite{smith1} is a well known concept describing the stable
states of a population resulting from dynamics of evolution. ESSs are known
to be symmetric Nash equilibria robust against small mutations \cite{damme}.
We explored how a strategy, being an ESS in classical version of the game,
performs if the game is played in quantum settings. Playing a game in
Marinatto and Weber's scheme \cite{marinatto} with particular choice of
initial quantum state, i.e., $\left| \psi _{ini}\right\rangle =c_{11}\left|
1,1\right\rangle +c_{22}\left| 2,2\right\rangle $, where $1$ and $2$
represent the classical pure strategies and $\left| c_{11}\right|
^{2}+\left| c_{22}\right| ^{2}=1$ with $c_{11},c_{22}\in \mathbf{C}$, we
showed that evolutionary stability of a pure strategy in a symmetric quantum
form of a bi-matrix game can be changed by a control on the parameters of
the initial quantum state \cite{iqbal,iqbal1}. However, with an initial
state in this form, the evolutionary stability of a mixed NE cannot be
changed for two-player games but it becomes possible when the number of
players is increased from two to three \cite{iqbal2}. In these
considerations the corresponding symmetric NE remain intact both in the
quantum and classical versions of the game.

In evolutionary game theory mixed strategies play a significant role. The
well-known \textit{Bishop-Cannings theorem} (BCT) \cite{bishop} describes an
interesting property of mixed ESSs in symmetric bi-matrix games. It is
useful to introduce the concept of \textit{support }of an ESS to understand
more easily the BCT \cite{vickers,mark}. Suppose a strategy vector $\mathbf{%
p=(}p_{i}\mathbf{)}$ is an ESS. Its support $S(\mathbf{p})$ is the set $S(%
\mathbf{p})=\left\{ i:p_{i}>0\right\} $. Thus the support of $\mathbf{p}$ is
the set of pure strategies that can be played by a $\mathbf{p}$-player. BCT
states that if $\mathbf{p}$ is an ESS with support I\textbf{\ }and\textbf{\ }%
$\mathbf{r}$\textbf{\ }$\neq \mathbf{p}$ is an ESS with support $J$, then $I$
$\nsupseteq $ $J$. For bi-matrix games the BCT\ shows that \textit{no pure
strategy can be evolutionary stable when a mixed ESS exists} \cite{mark}.
Naturally one, then, asks about the classical pure ESSs when a switch-over
to a quantum form of a classical symmetric bi-matrix game also gives
evolutionary stability to a mixed symmetric NE.

In present paper, following an approach developed for the quantum version of
the rock-scissor-paper (RSP) game \cite{iqbal3}, we consider a general form
of a two-qubit initial quantum state. Our results show that for this form of
initial quantum state, the corresponding quantum version of a bi-matrix game
can give evolutionary stability to a mixed NE, when classically it is not
stable. It is interesting to observe that by ensuring evolutionary stability
to a mixed NE in a quantum form of the game, the BCT forces out the pure
ESSs present in classical form of the game.

\section{Evolutionary stability of a mixed NE}

In a classical symmetric bi-matrix game, played in an evolutionary set-up
involving a population, all the members of the population are
indistinguishable and each individual is equally likely to face each other.
In such a set-up one assumes that individuals interact only in pair-wise
encounters. Suppose that the finite set of pure strategies $\left\{
1,2,...,n\right\} $ is available to each player. In one pair-wise encounter
let a player $A$ receives a reward $a_{ij}$ by playing strategy $i$ against
another player $B$ playing strategy $j$. In symmetric situation the player $%
B $, then, gets $a_{ji}$ as a reward. The value $a_{ij}$ is an element in
the $n\times n$ payoff matrix $\mathbf{M}$. We assume that the players also
have an option to play a mixed strategy. It means he/she plays the strategy $%
i$ with probability $p_{i}$ for all $i=1,2,...,n.$ A strategy vector $%
\mathbf{p,}$ with components $p_{i},$ represents the mixed strategy played
by the player. In standard notation an average, or expected, payoff for
player $A,$ playing strategy $\mathbf{p,}$ against player $B$ playing $%
\mathbf{q,}$ is written as $P(\mathbf{p,q)}$ \cite{mark}

\begin{equation}
P(\mathbf{p,q)=}\sum a_{ij}p_{i}q_{j}=\mathbf{p}^{T}\mathbf{Mq}
\label{payoffs}
\end{equation}
where $T$ is for transpose. Suppose that the strategy $\mathbf{p}$ is played
by almost all the members of the population, the rest of population forms a
small mutant group constituting a fraction $\epsilon $ of the total
population playing $\mathbf{q.}$ $\mathbf{p}$ is said to be \textit{%
evolutionary stable} against $\mathbf{q}$ if

\begin{equation}
P\left[ \mathbf{p,(}1-\epsilon )\mathbf{p+}\epsilon \mathbf{q}\right] >P%
\left[ \mathbf{q,(}1-\epsilon )\mathbf{p+}\epsilon \mathbf{q}\right] 
\label{ESSCond}
\end{equation}
for all sufficiently small $\epsilon $. Thus $\mathbf{p}$ does better
against the mean population strategy than $\mathbf{q}$ does. The condition (%
\ref{ESSCond}) implies that either (\textit{i}) $P(\mathbf{p,p)>}P(\mathbf{q}%
,\mathbf{p})$ or (\textit{ii}) $P(\mathbf{p,p)=}P(\mathbf{q},\mathbf{p})$
and $P(\mathbf{p,q)>}P(\mathbf{q},\mathbf{q})$. The vector $\mathbf{p}$ is
said to be an ESS if $\mathbf{p}$ is evolutionary stable against all $%
\mathbf{q}\neq \mathbf{p}.$

The payoff to a player in the quantum version of rock-scissors-paper (RSP)
game \cite{iqbal3} can also be written in similar form to (\ref{payoffs}),
provided the matrix $\mathbf{M}$ is replaced with a matrix corresponding to
the quantum version of the game. In RSP each player has access to three pure
strategies, represented by $1,2,$ and $3,$ and the game is given by the
following matrix, with the players recognized as Alice and Bob

\begin{equation}
\text{Alice}^{\prime }\text{s strategy \ }\overset{\text{Bob's strategy}}{
\begin{array}{c}
1 \\ 
2 \\ 
3
\end{array}
\overset{
\begin{array}{ccccccccc}
1 &  &  &  & 2 &  &  &  & 3
\end{array}
}{\left( 
\begin{array}{ccc}
(\alpha _{11},\alpha _{11}) & (\alpha _{12},\alpha _{21}) & (\alpha
_{13},\alpha _{31}) \\ 
(\alpha _{21},\alpha _{12}) & (\alpha _{22},\alpha _{22}) & (\alpha
_{23},\alpha _{32}) \\ 
(\alpha _{31},\alpha _{13}) & (\alpha _{32},\alpha _{23}) & (\alpha
_{33},\alpha _{33})
\end{array}
\right) }}  \label{cmatrix}
\end{equation}
where, for example, $(\alpha _{23},\alpha _{32})$\emph{\ }means that\emph{\ }%
Alice and Bob get $\alpha _{23}$ and $\alpha _{32}$, respectively, when
Alice plays the strategy $2$ and Bob plays $3$.\emph{\ }In quantum version
of the game the players apply unitary operators $I,C,$ and $D$ on an initial
quantum state defined as follows \cite{marinatto,iqbal3}:

\begin{eqnarray}
I\left| 1\right\rangle &=&\left| 1\right\rangle \qquad C\left|
1\right\rangle =\left| 3\right\rangle \qquad D\left| 1\right\rangle =\left|
2\right\rangle  \notag \\
I\left| 2\right\rangle &=&\left| 2\right\rangle \qquad C\left|
2\right\rangle =\left| 2\right\rangle \qquad D\left| 2\right\rangle =\left|
1\right\rangle  \notag \\
I\left| 3\right\rangle &=&\left| 3\right\rangle \qquad C\left|
3\right\rangle =\left| 1\right\rangle \qquad D\left| 3\right\rangle =\left|
3\right\rangle
\end{eqnarray}
where $C^{\dagger }=C=C^{-1}$ and $D^{\dagger }=D=D^{-1}$ and $I$ is the
identity operator. Suppose Alice applies the operators $C$, $D$, and $I$
with the probabilities $p,$ $p_{1\text{,}}$ and $(1-p-p_{1})$, respectively.
Similarly Bob applies the operators $C$, $D$, and $I$ with probabilities $q$%
, $q_{1}$, and $(1-q-q_{1})$ respectively, on the initial quantum state $%
\left| \psi _{ini}\right\rangle $ where

\begin{equation}
\left| \psi _{ini}\right\rangle =\underset{i,j=1,2,3}{\sum }c_{ij}\left|
i,j\right\rangle \text{ \ \ where \ \ }\underset{i,j=1,2,3}{\sum }\left|
c_{ij}\right| ^{2}=1  \label{inistat}
\end{equation}
The payoff to Alice who plays the strategy $\mathbf{p}$ (where $\mathbf{p}%
^{T}\mathbf{=}[1-p-p_{1},$ \ \ $p_{1},$ \ \ $p]$) against Bob who plays the
strategy $\mathbf{q}$ (where $\mathbf{q}^{T}\mathbf{=}[1-q-q_{1},$ \ $q_{1},$
\ \ $q]$) can be written as \cite{iqbal3}

\begin{equation}
P_{A}(\mathbf{p,q)=p}^{T}\mathbf{\omega q}
\end{equation}
where the matrix $\mathbf{\omega }$ is given by

\begin{equation}
\mathbf{\omega =}\left( 
\begin{array}{ccc}
\omega _{11} & \omega _{12} & \omega _{13} \\ 
\omega _{21} & \omega _{22} & \omega _{23} \\ 
\omega _{31} & \omega _{32} & \omega _{33}
\end{array}
\right)  \label{qmatrix}
\end{equation}
and the elements of $\mathbf{\omega }$ are given by following matrix equation

\begin{eqnarray}
&&\left( 
\begin{array}{ccccccccc}
\omega _{11} & \omega _{12} & \omega _{13} & \omega _{21} & \omega _{22} & 
\omega _{23} & \omega _{31} & \omega _{32} & \omega _{33}
\end{array}
\right)  \notag \\
&=&\left( 
\begin{array}{ccccccccc}
\alpha _{11} & \alpha _{12} & \alpha _{13} & \alpha _{21} & \alpha _{22} & 
\alpha _{23} & \alpha _{31} & \alpha _{32} & \alpha _{33}
\end{array}
\right) \times  \notag \\
&&\left( 
\begin{array}{ccccccccc}
\left| c_{11}\right| ^{2} & \left| c_{12}\right| ^{2} & \left| c_{13}\right|
^{2} & \left| c_{21}\right| ^{2} & \left| c_{22}\right| ^{2} & \left|
c_{23}\right| ^{2} & \left| c_{31}\right| ^{2} & \left| c_{32}\right| ^{2} & 
\left| c_{33}\right| ^{2} \\ 
\left| c_{12}\right| ^{2} & \left| c_{11}\right| ^{2} & \left| c_{12}\right|
^{2} & \left| c_{22}\right| ^{2} & \left| c_{21}\right| ^{2} & \left|
c_{22}\right| ^{2} & \left| c_{32}\right| ^{2} & \left| c_{31}\right| ^{2} & 
\left| c_{32}\right| ^{2} \\ 
\left| c_{13}\right| ^{2} & \left| c_{13}\right| ^{2} & \left| c_{11}\right|
^{2} & \left| c_{23}\right| ^{2} & \left| c_{23}\right| ^{2} & \left|
c_{21}\right| ^{2} & \left| c_{33}\right| ^{2} & \left| c_{33}\right| ^{2} & 
\left| c_{31}\right| ^{2} \\ 
\left| c_{21}\right| ^{2} & \left| c_{22}\right| ^{2} & \left| c_{23}\right|
^{2} & \left| c_{11}\right| ^{2} & \left| c_{12}\right| ^{2} & \left|
c_{13}\right| ^{2} & \left| c_{21}\right| ^{2} & \left| c_{22}\right| ^{2} & 
\left| c_{23}\right| ^{2} \\ 
\left| c_{22}\right| ^{2} & \left| c_{21}\right| ^{2} & \left| c_{22}\right|
^{2} & \left| c_{12}\right| ^{2} & \left| c_{11}\right| ^{2} & \left|
c_{12}\right| ^{2} & \left| c_{22}\right| ^{2} & \left| c_{21}\right| ^{2} & 
\left| c_{22}\right| ^{2} \\ 
\left| c_{23}\right| ^{2} & \left| c_{23}\right| ^{2} & \left| c_{21}\right|
^{2} & \left| c_{13}\right| ^{2} & \left| c_{13}\right| ^{2} & \left|
c_{11}\right| ^{2} & \left| c_{23}\right| ^{2} & \left| c_{23}\right| ^{2} & 
\left| c_{21}\right| ^{2} \\ 
\left| c_{31}\right| ^{2} & \left| c_{32}\right| ^{2} & \left| c_{33}\right|
^{2} & \left| c_{31}\right| ^{2} & \left| c_{32}\right| ^{2} & \left|
c_{33}\right| ^{2} & \left| c_{11}\right| ^{2} & \left| c_{12}\right| ^{2} & 
\left| c_{13}\right| ^{2} \\ 
\left| c_{32}\right| ^{2} & \left| c_{31}\right| ^{2} & \left| c_{32}\right|
^{2} & \left| c_{32}\right| ^{2} & \left| c_{31}\right| ^{2} & \left|
c_{32}\right| ^{2} & \left| c_{12}\right| ^{2} & \left| c_{11}\right| ^{2} & 
\left| c_{12}\right| ^{2} \\ 
\left| c_{33}\right| ^{2} & \left| c_{33}\right| ^{2} & \left| c_{31}\right|
^{2} & \left| c_{33}\right| ^{2} & \left| c_{33}\right| ^{2} & \left|
c_{31}\right| ^{2} & \left| c_{13}\right| ^{2} & \left| c_{13}\right| ^{2} & 
\left| c_{11}\right| ^{2}
\end{array}
\right)  \notag \\
&&  \label{qmatrix2}
\end{eqnarray}

The above matrix (\ref{qmatrix}) reduces to its classical form of Eq.(\ref
{cmatrix}), by making the initial state unentangled i.e., $\left|
c_{11}\right| ^{2}=1$.

In a symmetric game the exchange of strategies by Alice and Bob also
exchanges their respective payoffs. The concept of an ESS was originally
defined for symmetric games where a player's payoff is given by his strategy
and his identity does not affect it \cite{smith1,weibull}. It is seen that
the quantum game corresponding to the matrix (\ref{cmatrix}), when played
using the initial quantum state of eq. (\ref{inistat}), becomes symmetric
when

\begin{equation}
\left| c_{ij}\right| ^{2}=\left| c_{ji}\right| ^{2}\text{ for }i\neq j
\label{SymCond}
\end{equation}
Here the two-player quantum game, with three pure strategies, has a form
similar to a classical matrix game. The payoff matrix of the classical game
is, however, replaced now with its quantum version (\ref{qmatrix}). Also the
matrix (\ref{qmatrix}) now involves the coefficients $c_{ij}$ of the initial
quantum state (\ref{inistat}).

To reduce the above mathematical formalism to two-players, two-strategy
quantum game let us fix $p_{1}=q_{1}=0,$ i.e., both players do not use the
operator $D$ at all, and apply only the operators $C$ and $I$, with
classical probabilities, on the initial quantum state. Payoff to the player
who plays the strategy vector $\mathbf{p}$ (where $\mathbf{p}^{T}\mathbf{=}%
[1-p$ \ \ $p]$) against the player playing the strategy vector $\mathbf{q}$
(where $\mathbf{q}^{T}\mathbf{=}[1-q$ \ \ $q]$) can again be written as $P(%
\mathbf{p,q)=p}^{T}\mathbf{\omega q}$. Nevertheless, $\mathbf{\omega }$ is
now reduced to its simpler form given as 
\begin{equation}
\mathbf{\omega }=\left( 
\begin{array}{cc}
\omega _{11} & \omega _{13} \\ 
\omega _{31} & \omega _{33}
\end{array}
\right)
\end{equation}
where the elements of the matrix are

\begin{equation}
\left( 
\begin{array}{c}
\omega _{11} \\ 
\omega _{13} \\ 
\omega _{31} \\ 
\omega _{33}
\end{array}
\right) =\left( 
\begin{array}{cccc}
\left| c_{11}\right| ^{2} & \left| c_{13}\right| ^{2} & \left| c_{31}\right|
^{2} & \left| c_{33}\right| ^{2} \\ 
\left| c_{13}\right| ^{2} & \left| c_{11}\right| ^{2} & \left| c_{33}\right|
^{2} & \left| c_{31}\right| ^{2} \\ 
\left| c_{31}\right| ^{2} & \left| c_{33}\right| ^{2} & \left| c_{11}\right|
^{2} & \left| c_{13}\right| ^{2} \\ 
\left| c_{33}\right| ^{2} & \left| c_{31}\right| ^{2} & \left| c_{13}\right|
^{2} & \left| c_{11}\right| ^{2}
\end{array}
\right) \left( 
\begin{array}{c}
\alpha _{11} \\ 
\alpha _{13} \\ 
\alpha _{31} \\ 
\alpha _{33}
\end{array}
\right)  \label{terms}
\end{equation}
It is now a bi-matrix game played with the initial quantum state (\ref
{inistat}). The available pure strategies are now $1$ and $3$ only and the
terms with subscripts containing $2$ disappear. Take $x=1-p$ and $y=1-q,$ so
that $x$ and $y$ are probabilities with which players apply identity
operator on the initial state $\left| \psi _{ini}\right\rangle $. The
strategy vectors $\mathbf{p}$ and $\mathbf{q}$ can then be represented only
by the numbers $x$ and $y$, respectively. Payoff to a $x$-player against a $%
y $-player is obtained as

\begin{equation}
P(x\mathbf{,}y\mathbf{)=p}^{T}\mathbf{\omega q=}x\left\{ \omega
_{11}y+\omega _{13}(1-y)\right\} +(1-x)\left\{ \omega _{31}y+\omega
_{33}(1-y)\right\} .
\end{equation}
Suppose $(x^{\star },x^{\star })$ is a Nash equilibrium, i.e.,

\begin{eqnarray}
&&P(x^{\star },x^{\star })-P(x,x^{\star })  \notag \\
&=&(x^{\star }-x)\left\{ x^{\star }(\omega _{11}-\omega _{13}-\omega
_{31}+\omega _{33})+(\omega _{13}-\omega _{33})\right\} \geq 0
\end{eqnarray}
for all $x\in \lbrack 0,1]$. The mixed strategy $x^{\star }=x_{q}^{\star }=%
\frac{\omega _{33}-\omega _{13}}{\omega _{11}-\omega _{13}-\omega
_{31}+\omega _{33}}$ makes the payoff difference $P(x^{\star },x^{\star
})-P(x,x^{\star })$ identically zero. The subscript $q$ is for `quantum'.
Let $\bigtriangleup x=x^{\star }-x$ then

\begin{equation}
P(x_{q}^{\star },x)-P(x,x)=-(\bigtriangleup x)^{2}\left\{ \omega
_{11}-\omega _{13}-\omega _{31}+\omega _{33}\right\}  \label{ESS2}
\end{equation}
Now $x_{q}^{\star }$ is an ESS if $\left\{ P(x_{q}^{\star
},x)-P(x,x)\right\} >0$ for all $x\neq x_{q}^{\star }$ \cite{iqbal,iqbal1},
which leads to the requirement $(\omega _{11}-\omega _{31}-\omega
_{13}+\omega _{33})<0.$

The classical game corresponds when $\left| c_{11}\right| ^{2}=1$ and it
gives $\omega _{11}=\alpha _{11}$, $\omega _{13}=\alpha _{13},$ $\omega
_{31}=\alpha _{31}$, and $\omega _{33}=\alpha _{33}$, in accordance with the
Eq. (\ref{qmatrix2})$.$ In case $(\alpha _{11}-\alpha _{13}-\alpha
_{31}+\alpha _{33})>0$, the mixed NE of a classical game, i.e., $x^{\star
}=x_{c}^{\star }=\frac{\alpha _{33}-\alpha _{13}}{\alpha _{11}-\alpha
_{13}-\alpha _{31}+\alpha _{33}}$ is not an ESS. Here the subscript $c$ is
for `classical'. Since we are interested in a situation where evolutionary
stability of a symmetric NE changes --while the corresponding NE remains
intact-- in transforming the game from classical to quantum form, lets take

\begin{equation}
x_{c}^{\star }=x_{q}^{\star }=\frac{\alpha _{33}-\alpha _{13}}{\alpha
_{11}-\alpha _{31}-\alpha _{13}+\alpha _{33}}=\frac{\omega _{33}-\omega _{13}%
}{\omega _{11}-\omega _{31}-\omega _{13}+\omega _{33}}.  \label{mNE1}
\end{equation}
saying that the classical NE $x_{c}^{\star }$ is also a NE in quantum form
of the game. One notices from the matrix in the Eq. (\ref{terms})

\begin{eqnarray}
&&(\omega _{11}-\omega _{31}-\omega _{13}+\omega _{33})  \notag \\
&=&(\alpha _{11}-\alpha _{13}-\alpha _{31}+\alpha _{33})(\left|
c_{11}\right| ^{2}-\left| c_{13}\right| ^{2}-\left| c_{31}\right|
^{2}+\left| c_{33}\right| ^{2})  \label{mNE2}
\end{eqnarray}
and 
\begin{eqnarray}
\omega _{33}-\omega _{13} &=&\left| c_{11}\right| ^{2}(\alpha _{33}-\alpha
_{13})+\left| c_{13}\right| ^{2}(\alpha _{31}-\alpha _{11})+  \notag \\
&&\left| c_{31}\right| ^{2}(\alpha _{13}-\alpha _{33})+\left| c_{33}\right|
^{2}(\alpha _{11}-\alpha _{31})  \label{mNE3}
\end{eqnarray}
Now a substitution from Eqs. (\ref{mNE2},\ref{mNE3}) into the Eq. (\ref{mNE1}%
) gives $\alpha _{33}-\alpha _{13}=\alpha _{11}-\alpha _{31}$, and this
leads to $x_{c}^{\star }=x_{q}^{\star }=\frac{1}{2}$. Therefore, the mixed
strategy $x^{\star }=\frac{1}{2},$ remain a NE in both classical and a
quantum form of the game. Consider now this mixed NE for a classical game
with $(\alpha _{11}-\alpha _{13}-\alpha _{31}+\alpha _{33})>0$ -- showing
that it is not an ESS. The above Eq. (\ref{mNE2}) shows an interesting
possibility that it is still possible to have $(\omega _{11}-\omega
_{31}-\omega _{13}+\omega _{33})<0$ if

\begin{equation}
(\left| c_{11}\right| ^{2}+\left| c_{33}\right| ^{2})<(\left| c_{13}\right|
^{2}+\left| c_{31}\right| ^{2})  \label{IniCond}
\end{equation}
In other words, now the evolutionary stability of a mixed strategy --which
is a NE in both classical and quantum versions of the game-- changes when
the game switches-over between its two forms. To have a symmetric game in
its quantum form one also needs $\left| c_{13}\right| ^{2}=\left|
c_{31}\right| ^{2}$ and the inequality (\ref{IniCond}) reduces to $\left|
c_{11}\right| ^{2}+\left| c_{33}\right| ^{2}<\left| c_{13}\right|
^{2}+\left| c_{31}\right| ^{2}$.

Therefore, a quantum version of a symmetric bi-matrix classical game of the
matrix

\begin{equation}
\left( 
\begin{array}{cc}
(\alpha _{11},\alpha _{11}) & (\alpha _{13},\alpha _{31}) \\ 
(\alpha _{31},\alpha _{13}) & (\alpha _{33},\alpha _{33})
\end{array}
\right)
\end{equation}
can be played by players having two unitary operators and a general
two-qubit quantum state of the form

\begin{equation}
\left| \psi _{ini}\right\rangle =\underset{i,j=1,3}{\sum }c_{ij}\left|
ij\right\rangle  \label{SymmInit}
\end{equation}
where $\underset{i,j=1,3}{\sum }\left| c_{ij}\right| ^{2}=1$. In case $%
\alpha _{33}-\alpha _{13}=\alpha _{11}-\alpha _{31}$ the mixed strategy $%
x^{\star }=\frac{1}{2}$ is not an ESS in the classical game if $(\alpha
_{33}-\alpha _{13})>0$. Nevertheless, the strategy $x^{\star }=\frac{1}{2}$
becomes an ESS when $\left| c_{11}\right| ^{2}+\left| c_{33}\right|
^{2}<\left| c_{13}\right| ^{2}+\left| c_{31}\right| ^{2}$. In case $(\alpha
_{33}-\alpha _{13})<0$ the strategy $x^{\star }=\frac{1}{2}$ is an ESS
classically but does not remain if $\left| c_{11}\right| ^{2}+\left|
c_{33}\right| ^{2}<\left| c_{13}\right| ^{2}+\left| c_{31}\right| ^{2}$. Now
suppose $\left| c_{13}\right| ^{2}=\left| c_{31}\right| ^{2}=0$. Then the
Eq. (\ref{mNE2}) reduces to

\begin{equation}
(\omega _{11}-\omega _{13}-\omega _{31}+\omega _{33})=(\alpha _{11}-\alpha
_{13}-\alpha _{31}+\alpha _{33})
\end{equation}
One observes from the above equation that if a quantum game is played by
following simple form of the initial quantum state

\begin{equation}
\left| \psi _{ini}\right\rangle =c_{11}\left| 11\right\rangle +c_{33}\left|
33\right\rangle   \label{SimpInit}
\end{equation}
it is not possible to influence the evolutionary stability of a mixed NE, as
it is concluded in our earlier work \cite{iqbal1,iqbal2}.

\section{Summary}

Mixed ESSs appear in many games of interest that are played in the natural
world. The examples of the Rock-Scissors-Paper (RSP) and the Hawks and Doves
games are well known from evolutionary game theory. In evolutionary game
theory the Bishop-Cannings theorem does not permit pure ESSs when a mixed
ESS exists in a bi-matrix game. In earlier work \cite
{iqbal,iqbal1,iqbal2,iqbal3} we showed that it is possible to change
evolutionary stability of a pure symmetric NE with a control of the
parameters $c_{11}$ and $c_{22}$ when the game is played with an initial
two-qubit quantum state of the form $\left| \psi _{ini}\right\rangle
=c_{11}\left| 1,1\right\rangle +c_{22}\left| 2,2\right\rangle $ where $%
\left| c_{11}\right| ^{2}+\left| c_{22}\right| ^{2}=1$. However,
evolutionary stability of a mixed symmetric NE cannot be changed with such a
control. In this paper, following the approach developed for the quantum
version of the rock-scissor-paper (RSP) game \cite{iqbal3}, we allowed the
game to be played with a general form of a two-qubit initial quantum state.
With this state it becomes possible to change evolutionary stability of a
mixed NE. For a bi-matrix game we worked out a symmetric mixed NE that
remains intact in both the classical and quantum versions of the game. For
this mixed NE we, then, found conditions making it possible that
evolutionary stability of a mixed symmetric NE changes with a switch-over of
the game between its two forms, one classical and the other quantum.

\end{document}